\documentclass{INTERSPEECH2023}

\interspeechcameraready

\DeclareMathOperator*{\argmin}{arg\,min}
\usepackage{adjustbox}
\usepackage[dvipsnames]{xcolor}
\usepackage{amsmath}
\usepackage{algorithm}
\usepackage{algpseudocode}

\title{Automatic Data Augmentation for Domain Adapted Fine-Tuning of Self-Supervised Speech Representations}
\name{Salah Zaiem$^1$, Titouan Parcollet$^{2,3}$, Slim Essid$^1$}
\address{
  $^1$LTCI, Télécom Paris, Institut Polytechnique de Paris, France\\
  $^2$Samsung AI Center, Cambridge, United-Kingdom\\
  $^3$University of Cambridge, United-Kingdom}
\email{salah.zaiem@telecom-paris.fr}

\begin{document}

\maketitle
\begin{abstract}
Self-Supervised Learning (SSL) has allowed leveraging large amounts of unlabeled speech data to improve the performance of speech recognition models even with small annotated datasets. Despite this, speech SSL representations may fail while facing an acoustic mismatch between the pretraining and target datasets. To address this issue, we propose a novel supervised domain adaptation method, designed for cases exhibiting such a mismatch in acoustic domains. It consists in applying properly calibrated data augmentations on a large clean dataset, bringing it closer to the target domain, and using it as part of an initial fine-tuning stage. Augmentations are automatically selected through the minimization of a conditional-dependence estimator, based on the target dataset. The approach is validated during an oracle experiment with controlled distortions and on two amateur-collected low-resource domains, reaching better performances compared to the baselines in both cases.
\end{abstract}

\section{Introduction}

Self-supervised learning (SSL) enables the use of large amounts of unlabelled data to obtain substantial performance improvements in a variety of downstream tasks without relying on manual annotations. Various approaches have been introduced including predictive coding \cite{baevski2020wav2vec,Liu_2020}, multi-task learning \cite{ravanelli2020multitask, zaiem2022pretext}, auto-encoding techniques \cite{algayres:hal-02977539} or contrastive learning \cite{cola, Shor_2020}. In this context, data augmentation has become an important part of many self-supervised approaches. Particularly, various studies have shown that applying several distortions during pretraining leads to more robust representations, either with  Contrastive Predictive Coding (CPC) \cite{kharitonov}, or with Wav2vec2.0 \cite{Sriram2022, Riviere2021}. Recently, WavLM \cite{Chen2021} incorporated distortions to add a denoising criterion to its predictive objective.

However, and despite its success, self-supervised learning has been shown to suffer from domain mismatch where the fine-tuning samples from the target domain are vastly different from the pretraining ones \cite{Hsu2021, Riviere2021}. While progress has been made in achieving near-optimal performance on clean datasets such as LibriSpeech, spontaneous speech datasets and non-professionally recorded ones still exhibit lower performance, as displayed in recent speech SSL benchmarks \cite{tsai-etal-2022-superb, evain2021lebenchmark}.

To mitigate the performance drop caused by domain mismatch, various domain adaptation techniques have been explored, particularly in transfer learning settings \cite{Olvera2022}. In the self-supervised context, adversarial approaches have been applied during the unsupervised pretraining and tested on speech recognition \cite{Tanaka2022, lodagala2023pada}, emotion recognition \cite{Latif2022} and speaker recognition \cite{chen2021self}. Along with domain adversarial paradigms, Huang \textit{and al.} \cite{huang2022} investigated continual learning methods during pretraining. Distinctly, our method does not aim at aligning latent representations, but rather transforms the audio waveforms of a neutral dataset to match the acoustic conditions of the target domain using data augmentations, rendering this dataset better suited to the final task in an initial fine-tuning stage.

Furthermore, retraining the self-supervised feature extractors with additional domain-invariant enforcement, as proposed in the literature, is a hard and costly endeavor, with the latest SSL models being trained on $94k$ hours of audio data using 64 V100 GPUs \cite{Chen2021}. Thus, we envisage the option of augmenting a supposedly neutral dataset and using it for the first fine-tuning step. The augmentations to be applied and their parameters are chosen in order to optimize the similarity in terms of recording conditions between the modified and the target dataset and hence the final performance. Our method presents three main advantages. First, it enables the use of large and clean available annotated datasets, enhancing the textual diversity of the training corpus. Second, it does not require a new pretraining as it directly fine-tunes available SSL models. Finally, it allows an efficient data augmentation exploration, as the selection and parametrization is automatic and does not involve any neural network training. It is, thus, largely more efficient than thorough testing, as scoring $200$ augmentation policies takes $3$ hours on $10$ CPUs, while complete testing of one augmentations distribution necessitates around $20$ hours of GPU computations. 

The contributions of this work are two-fold: i) Propose a new method for supervised domain adaptation consisting in applying appropriate signal distortions to a clean labeled dataset used for an initial fine-tuning step. The method is validated with an oracle simulated experiment and experiments with naturally noisy datasets. ii) Release the code base, implemented with SpeechBrain \cite{speechbrain} for replication and further improvements.\footnote{\url{https://github.com/salah-zaiem/augmentations_adaptation}}

 \begin{figure*}[ht!]
  \centering
  \includegraphics[width=0.6\linewidth]{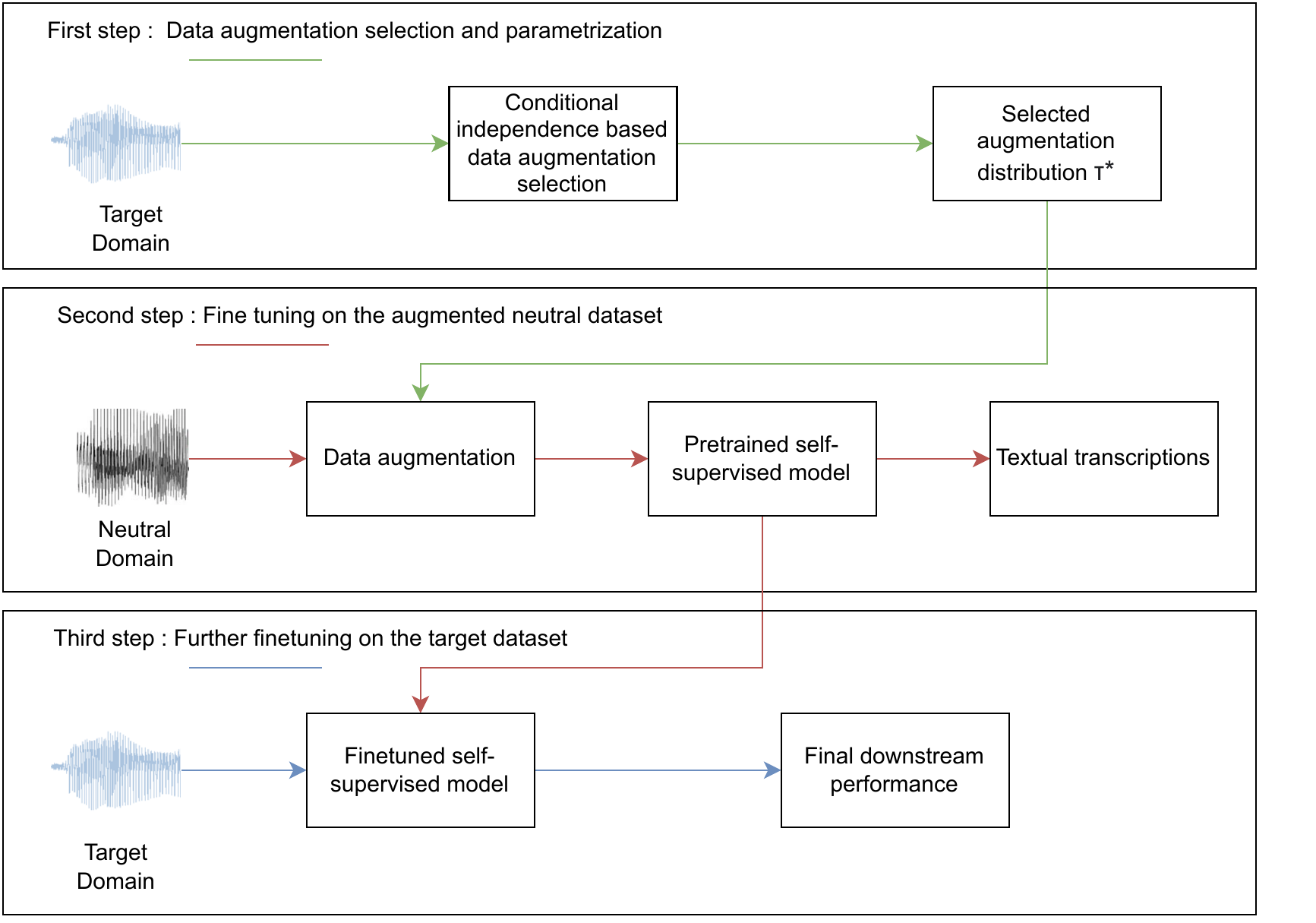}
  \caption{Summary of the three steps of the method. 1. Starting from the target domain, an augmentation distribution is computed. 2. This distribution is used to distort a neutral dataset for a first fine-tuning. 3. A final fine-tuning is done on the target domain samples.}
  \label{fig:schema}
\vspace{-0.5cm}
\end{figure*}

Figure \ref{fig:schema} presents an overview of the method, summarizing the three steps conducted for every considered target dataset. First, and given the labeled target dataset, an augmentation distribution is automatically selected (Section \ref{sec:selection}). Second, a first fine-tuning of the self-supervised representation is done, using the neutral dataset distorted with the augmentations selected in the first step. Finally, a second fine-tuning on the small target domain dataset is done leading to the final model that will be evaluated using the target test set (Section \ref{sec:natural}). Experiments show a speech recognition performance relative improvement reaching $19.5\%$ in a real-world distorted dataset scenario. 

\section{Selecting the Augmentation Distribution}
\label{sec:selection}
Given a labeled target speech recognition dataset, our method selects an augmentation distribution that is best suited to its recording conditions. From this distribution,  we will sample augmentations to be applied to a larger ``clean'' dataset which will be used to fine-tune the SSL representations. The goal is to select augmentations bringing the ``clean'' dataset samples ``closer'' to those of the target domain, thus leading to better performance on its test sets. This section details the conditional-independence-based method developed to select a data augmentation distribution given the annotated target dataset. It starts by detailing the motivations behind the method, before delving into the technical details of the implementation. 

\subsection{Motivation and Technical Description}
\label{subsec:motivation}
 \textbf{Motivation.} Inspired by pretext-tasks selection for speech self-supervised learning, Zaiem \textit{et al.} \cite{Zaiem2022} have shown that conditional independence estimation may be used for automatic data augmentation in contrastive self-supervised learning settings. Furthermore, qualitative analysis has indicated that the distortions selected by this technique tend to be close to those of the target downstream dataset. Explicitly, applying a set of augmentations creates a set of augmented versions, often called ``views", of the original samples. Minimizing, with respect to the augmentations selected, the dependence between the views and the IDs of the samples they originate from, conditionally on the downstream labels leads to a good choice of augmentations in contrastive learning settings.

 Let us give an intuition about what happens in these conditional independence computations to understand why it can be useful for domain adaptation as well. Roughly, minimizing the conditional dependence described above maximizes, within the same downstream class, the invariance of distorted samples (\textit{i.e.} views) to the ID of their original speech sample. If a given distortion (for instance, reverberation) is not present in any sample in the original target dataset, randomly applying this distortion would decrease in-class similarity. Inversely, applying augmentations already present in samples in the dataset makes it harder to distinguish their original samples' IDs given the distorted samples and, thus, lowers the conditional dependence estimator. Conditioning on the downstream labels retains the signal clues characterizing the downstream classes since it prevents selecting distortions that are only relevant to one class, as they would reduce in-class similarity in the other classes.

 \noindent\textbf{Technical Description.} Precisely, let $X$ and  $Y$ be respectively, a set of speech data points and their respective set of downstream labels which are in our case textual transcriptions. With $\tau$ an augmentation distribution from which one can sample a chain of augmentations, we compute a distorted dataset $X' = f(X, \tau)$, with $f$ a function that randomly applies augmentations sampled from $\tau$ on the speech samples. Specifically, we can generate $N$ augmented versions per speech sample to get the augmented set of data points $X'$, with $N$ a hyperparameter. Every sample $x'$ in $X'$ is a distorted version of a point $x$ in the original dataset $X$. We will refer to the ID of the original point $x$ as $z$, defining the $Z$ set. The ID here corresponds to a discrete value indexing the speech segments $X$. In contrastive self-supervised learning settings \cite{simclr}, augmentation selection is crucial to incorporate the most relevant invariances in the learned representation into the downstream task of interest \cite{Xiao2020}. In this context, it has been shown that choosing the augmentation distribution $\tau$ that minimizes an estimator of the conditional dependence between $X$ and $Z$ given $Y$ leads to the best downstream performance on speaker and language recognition tasks \cite{Zaiem2022}. This work extends this approach in two manners, first applying it for domain adaptation in a supervised setting, and second extending it to the speech recognition task. We use for this the Hilbert-Schmidt Independence Criterion (HSIC) \cite{gretton}, a kernel-based dependence estimator, also validated on pretext task selection in previous works \cite{https://doi.org/10.48550/arxiv.2107.00594}. The lower the HSIC estimator, the more conditionally independent the two sets are and the better the augmentations should be.

In summary, to find the optimal augmentation distribution $\tau^*$, we resort to minimizing the HSIC quantity with the augmented dataset $X' = f(X, \tau)$ according to $\tau^* = \argmin_\tau HSIC (f(X,\tau), Z | Y) $ 
\begin{table}[ht!]
  \caption{Augmentations, descriptions and parameter ranges}
  \label{tab:parameters}
  \centering
  \scalebox{0.80}{\begin{tabular}{lll}
    \toprule
    \multicolumn{1}{l}{\textbf{Name}} & 
                                         \multicolumn{1}{l}{\textbf{Description}}
                                     &    \multicolumn{1}{l}{\textbf{Range (Unit)}}\\
    \midrule
                              Low Min  &Lowpass minimal frequency cutoff              &  [100-500] (Hz) \\
     Low Max &Lowpass maximal frequency cutoff  & [1000-5000] (Hz)             \\

                              High Min   &Highpass minimal frequency cutoff               &  [1000,4000] (Hz) \\
     High Max  &Highpass maximal frequency cutoff  & [4000,6000] (Hz)             \\
     Pitch min  & Minimal pitch shift & [-6,-2] (semitones)             \\
     Pitch max  & Maximal pitch shift & [2,6] (semitones)             \\
     Min SNR  &Minimal SNR for coloured noise & [0,5] (dB)              \\
     Max SNR &Maximal SNR for coloured noise  & [10,30] (dB)        \\
     Min Gain & Minimal gain  & [-20,-10] (dB)\\
     Max Gain & Maximal gain  & [3,10] (dB)\\

    \bottomrule
  \end{tabular}}
  
\end{table}

with $HSIC (X', Z | Y)$ an estimate of the conditional dependence between the distorted speech samples and their original IDs given their downstream textual labels.

\subsection{Augmentation Distributions and Implementation}
An augmentation distribution $\tau$ is characterized by a set of parameters that defines how the chain of augmentations is sampled during training and applied to the next data point. Precisely, every distribution $(\tau(p))_{1\leq p \leq P}$ is represented as a vector of $P=17$ parameters representing either the probability of applying an augmentation or the boundaries of a uniform probability distribution used to sample the parameters of the augmentation (\textit{e.g.} maximal signal-to-noise ratio value for noise addition).

Since the considered augmentations are not differentiable according to the considered parameters, we apply a random search to minimize the HSIC value described above. Thus, we sample random distributions and select the one with the lowest dependence scoring. Specifically, for every considered target dataset, we first sample $D=100$ distribution parametrizations $(\tau_i)_{i \in [1,D]}$. For every parametrization $\tau_i$, we compute the HSIC quantity following two steps. First, the augmented set $X'_i = f(X, \tau_i)$ is generated by computing $N=20$ views of every speech sample in $X$. Then,  $HSIC(X_i', Z | Y)$ is computed following the technique described in \cite{Zaiem_2021}. For $Y$, we consider the $10$ classes consisting of the $10$ most used words in the dataset and take only the portion of the speech where the word is pronounced, using word-level forced alignment. The augmentation distribution with the lowest HSIC scoring is selected to be applied during fine-tuning.

\section{Experiments}
\label{experiments}

This section describes the experiments led to validate the proposed approach first in a simulated environment, then on real-world distorted datasets.

\subsection{Shared Experimental Protocol}

In all the experiments, the model is composed of two blocks: a pre-trained Wav2Vec2.0 Large model and a downstream decoder. The pre-trained model acts directly on the speech waveform and outputs an embedding of size $1,024$ every $20$ms of speech. Two fully connected layers with a hidden size of $1,024$ map each frame vector to one of the considered characters. The whole model is fine-tuned using Connectionist Temporal Classification (CTC) \cite{ctc} loss. During inference, greedy decoding is applied to the CTC probability outputs without any language-model-based re-scoring following the SpeechBrain recipe \cite{speechbrain}.  

We employ the Torch-Audiomentations library from the Asteroid team \cite{Pariente2020} as it accelerates the computation of augmentations both during HSIC scoring and training. From the pool of available augmentations, we selected the ones that have demonstrated efficacy in enhancing recognition performance with the contrastive predictive coding method \cite{kharitonov}. Hence, seven augmentations are considered: pitch shifting, reverberation, gain (which may reproduce clipping issues), colored noise addition, high and low pass filtering, and polarity inversion. The application of these distortions is controlled with a set of parameters listed in Table \ref{tab:parameters}.

\subsection{Oracle Experiment}

\textbf{Task-specific experimental protocol.} In this part, a known distortion distribution is first applied to a clean testing set. The resulting data will be considered as the mismatching target domain (i.e. a simulated one). In a second time, using this generated ``noisy" dataset, appropriate augmentations, selected using our conditional independence-based method, are applied to a clean training dataset that will be used for fine-tuning our self-supervised representations. As only the test set is distorted, this simulated experiment only involves one fine-tuning, contrarily to the real-data scenario, where a second fine-tuning stage is held on the target training data, as shown in Figure \ref{fig:schema}. This simulated experiment has two advantages compared to a natural setting. First, it ensures that the distortions in the testing set can be replicated by the set of augmentations considered. Second, since we have access to the augmentation distribution that generated the ``distorted" target dataset, it allows estimating the similarity between the augmentation distribution used to create the simulated testing domain and the one obtained with our method, \textit{i.e.} the similarity between the parameters controlling the chain of distortions applied.\\

In these experiments, $A=8$ augmentations distributions are sampled and applied on the LibriSpeech \textit{test-clean} and \textit{test-other} splits \cite{7178964}. For every sampled distribution, these two distorted splits are then considered as the testing datasets. We apply the same augmentation distributions to the \textit{dev-clean} and \textit{dev-other} splits, and use these two sets to compute the optimal augmentations following the method described in the previous section. Finally, we use the computed distribution $\tau^{*}$ with the lower $HSIC$ estimator value as the augmentation for fine-tuning our SSL model on LibriSpeech \textit{train-clean-100} split.

\begin{table}[]
\caption{Mean WER results on distorted versions of LibriSpeech test splits. While scoring below the topline, our method, named ``CI Augment", is significantly better than applying all or random augmentations. ``Baseline" corresponds to an augmentation-free training.}
  \label{tab:oracle_results}
  \centering
  \scalebox{0.75}{
\begin{tabular}{lcccc}
    \toprule

\multicolumn{1}{l}{\textbf{LS Split}} & \textbf{Baseline} & \textbf{Random} & \textbf{CI Augment} & {\color{olive}\textbf{Topline}} \\ 
\midrule
test-clean                                             & 29.86    & 29.91  & \textbf{27.20}  & {\color{olive} 26.11}   \\
test-other                                             & 43.89    & 42.48  & \textbf{40.68}  & {\color{olive}36.92}   \\
\bottomrule

\end{tabular}}
\vspace{-0.25cm}
\end{table}

\begin{table*}[]
\caption{Mean WER results on distorted versions of LibriSpeech test-clean and test-other. Our method, named ``CI Augment", outperforms the baselines and random augmentations for each one of the two contributors.}

\centering
  \scalebox{0.88}{

\begin{tabular}{lcccccc}
    \toprule
 \textbf{Contributor}        & \multicolumn{3}{c}{\textbf{Without Augmentations}}  & \multicolumn{3}{c}{\textbf{With Augmentations}}  \\
         \midrule

   & \textit{train-clean-100} & Contributor Only & \textit{train-clean-100} + Contributor & All        & Random      & CI Augment      \\
\midrule

Contributor 1 & 102.52   & 73.0        & 27.71             & 27.95      & 27.33       & \textbf{24.27}       \\
Contributor 2 & 96.49    & 98.92       & 20.48             & 20.76      & 22.23       & \textbf{16.49}      \\

\bottomrule

\end{tabular}}
\vspace{-0.35cm}

\label{tab:clients}

\end{table*}

\begin{figure}[ht!]

  \centering

  \includegraphics[width=0.70\linewidth, scale=0.2]{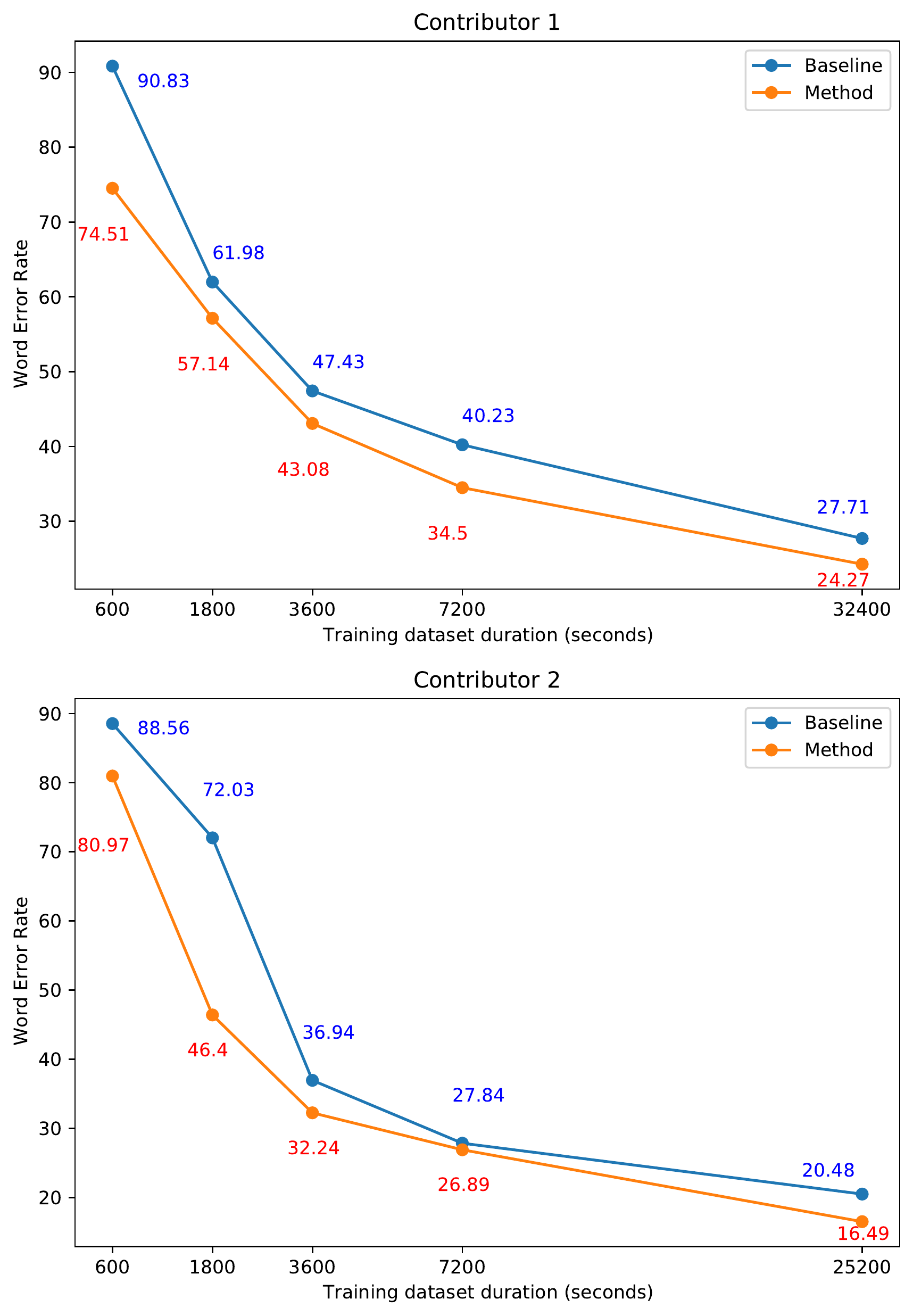}

  \vspace{-0.3cm}
  \caption{Effect of selecting augmentations on the performance depending on the quantity of target domain training data for each of the two considered contributors. The x-axis is not linear.}
  \label{fig:durations}

\vspace{-0.75cm}
\end{figure}
\noindent\textbf{Results.} Table \ref{tab:oracle_results} presents the results obtained on the test splits of LibriSpeech in the oracle experiments, with the column ``CI Augment" (the name of the approach, CI standing for Conditional Independence) showing the results of the proposed approach. Each value corresponds to the mean of the values obtained with each of the $A$ target datasets created with the sampled augmentation distributions. The ``Topline" corresponds to the result obtained when the training samples are augmented using the same distribution as the one used to generate the distorted testing splits (i.e oracle scenario). Two baselines are considered: the first one referred to as ``All" applies all the considered augmentations on the speech samples with their default parameters. Then, ``Random" refers to the mean value obtained if applying the $(A-1)$ randomly sampled augmentations, corresponding to the trainings already performed for the other toplines computation. Our method, while performing worse than the topline, leads to a relative word error rate (WER) improvement of $12.7\%$ compared to the baseline on \textit{test-clean}.

This controlled experiment also enables us to verify if the selected augmentations result in acoustic conditions cloning, as suggested in Section \ref{subsec:motivation}. Indeed, the probabilities of applying a given distortion to each testing set are known. To verify our intuition, for each one of the $8$ augmentation distributions applied, we sample $200$ other random augmentation distributions and score them using HSIC. For every scored distribution, we consider the vector composed by the seven probabilities of applying the considered distortions. Since these probabilities are known for the target distribution, we can compute an $L_2$ distance between the vector of probabilities of applying distortions used to create the target dataset, and those of the sampled scored distributions. We observe a Spearman correlation score of $0.51$ between the HSIC scores and the distances between vectors of probabilities. Furthermore, the application probabilities of the $10$ (top $5\%$) best scoring distributions are $15$\% closer to the target ones than those of the $10$ worst scoring ones. These results indicate that the selected augmentations, \textit{i.e.} those with low HSIC scoring, create samples closer to the target domain.

\subsection{Experiments with Naturally Distorted Datasets}\label{sec:natural}

In this section, we test and validate the proposed approach on  real low-resource ``noisy" datasets.\\ 

\noindent \textbf{Task-specific experimental protocol.} The goal is to adapt a large clean ``neutral" labeled dataset to better match the acoustic conditions of a small target dataset. The modified dataset is used during a first fine-tuning of the SSL representation, before further fine-tuning on the target dataset. To ensure a valid evaluation, the target dataset must meet two criteria: first, it should display consistent noisy recording and acoustic conditions. Second, neutral and target datasets should not exhibit different textual settings, \textit{i.e} differences such as spontaneous versus read speech, as our augmentations only address acoustic distortions. The Librispeech \textit{train-clean-100} is used as the clean dataset to be modified. The target datasets, on the other hand, correspond to the largest contributors of the CommonVoice 11.0 English dataset \cite{ardila2020common}. Starting from the ten most prolific contributors, two of them are finally selected after removing elements with heavy accents, and unintelligible or very clean recordings. For these two selected contributors, we partition the recorded samples into the train, validation, and test splits, and only use the training data to compute the augmentation distribution selection. The train splits are $9$ and $7$ hours long. More details can be found in the repository. \\

\noindent\textbf{Results.} Table \ref{tab:clients} reports the WERs with or without augmentations during the first fine-tuning on \textit{train-clean-100}. The first vertical part of the table shows the results obtained on the baselines without augmentations. ``train-clean-100" corresponds to fine-tuning only on Librispeech \textit{train-clean-100} split non-distorted. ``Contributor Only" corresponds to training only on the contributor data. For all other columns, the model is fine-tuned on \textit{train-clean-100} first, with or without augmentations, before further fine-tuning on the contributor data. The ``CI Augment" column shows that the augmentations chosen with our conditional-independence-based method lead to better target performance than applying no, all, or random augmentations on the neutral training split. The relative improvement compared to the augmentation-free baseline reaches $12.4\%$  for Contributor 1 and $19.5\%$ for Contributor 2.

Furthermore, we study how this affects the amount of target domain data needed (see Figure \ref{fig:durations}). We start by fine-tuning with the chosen distortions for the ``Method" lines and on the clean original LibriSpeech dataset for the ``Baseline" lines. Then, the duration of annotated target data used is augmented gradually . For the two contributors, the orange curve representing the evolution of the WER after fine-tuning with the computed distortions is always below the blue curve corresponding to the baseline. The effect is particularly visible with Contributor 1 with a performance $16.6$\% higher relatively when training with only $2$ hours.

\section{Acknowledgements}
This work has benefited from funding from l'Agence de l'Innovation de Défense, and was performed using HPC resources from GENCI-IDRIS (Grant 2021-AD011012801R1).

\section{Conclusion}
Self-supervised representations severely underperform when facing acoustic domain mismatch. We have introduced a method using automatic data augmentation selection to reduce the drop in performance when switching of acoustic domains. Experiments led in controlled and natural settings validate our assumption and method, and also show that it helps reduce the quantity of annotated data needed in the target domain. 

\bibliographystyle{IEEEtran}
\bibliography{mybib}

\end{document}